______________________________________________________________________________
\documentstyle[aps]{revtex}
\begin{document}
\title{Levinson's theorem for the Schr\"{o}dinger equation in one dimension}

\author{Shi-Hai Dong \thanks{Electronic
address:DONGSH@BEPC4.IHEP.AC.CN}}

\address{Institute of High Energy Physics, P.O. Box 918(4), Beijing 100039,
People's Republic of China}

\author{Zhong-Qi Ma}

\address{China Center for Advanced Science and Technology
(World Laboratory), P. O. Box 8730, Beijing 100080\\
and Institute of High Energy Physics, P. O. Box 918(4),
Beijing 100039, People's Republic of China}

\author{Martin Klaus}
\address{Department of Mathematics,
Virginia Polytechnic Institute and State University\\
Blacksburg, Virginia 24061, USA}


\maketitle

\vspace{4mm}

\begin{abstract}
Levinson's theorem for the one-dimensional Schr\"{o}dinger 
equation with a symmetric potential, which decays at 
infinity faster than $x^{-2}$, is established by the 
Sturm-Liouville theorem. The critical case, where the 
Schr\"{o}dinger equation has a finite zero-energy solution, 
is also analyzed. It is demonstrated that the number of bound 
states with even (odd) parity $n_{+}$ ($n_{-}$) is 
related to the phase shift $\eta_{+}(0)[\eta_{-}(0)]$ of the 
scattering states with the same parity at zero momentum as
$$\eta_{+}(0)+\pi/2=n_{+}\pi,~~~~~ \eta_{-}(0)=n_{-}\pi, ~~~~~ 
{\rm for~the~non-critical~case}, $$
$$\eta_{+}(0)=n_{+}\pi,~~~~~ \eta_{-}(0)-\pi/2=n_{-}\pi, ~~~~~ 
{\rm for~the~critical~case}, $$

\vskip 4mm
\noindent
PACS number(s): 03.65.Nk, 73.50.BK

\end{abstract}

\vskip 1cm
\section{INTRODUCTION}

The Levinson theorem [1], an important theorem 
in the nonrelativistic quantum scattering theory, 
established the relation between the total number $n_{\ell}$ 
of bound states with angular momentum $\ell$ and the phase 
shift $\delta_{\ell}(0)$ of the scattering state at zero 
momentum for the Schr\"{o}dinger equation with a spherically 
symmetric potential $V(r)$ in three dimensions:
$$\delta_{\ell}(0)-\delta_{\ell}(\infty)=\left\{\begin{array}{ll}
\left(n_{\ell}+1/2 \right)\pi~~~~~ &{\rm when}~~\ell=0 ~~{\rm and~
a~half-bound~state~occurs} \\
n_{\ell}\pi &{\rm the~remaining~cases}, \end{array} \right. 
 \eqno (1) $$

\noindent
where the potential $V(r)$ satisfies the following asymptotic conditions: 
$$r^{2}|V(r)| dr \longrightarrow  0, ~~~~~{\rm at}~~ r 
\longrightarrow  0, \eqno (2) $$
$$r^{3}|V(r)| dr \longrightarrow  0, ~~~~~{\rm at}~~ r
\longrightarrow \infty. \eqno (3) $$

\noindent
These two conditions are necessary for the nice behavior of 
the wave function at the origin and the analytic property of 
the Jost function, respectively. The first line in Eq. (1) was 
first expressed by Newton [2] for the case where a half-bound 
state of the $S$ wave occurs. The zero-energy solution to the
Schr\"{o}dinger equation is called a half-bound state provided 
that its wave function is finite, but does not decay fast enough 
at infinity to be square integrable. 

During the past half-century, the Levinson theorem has been 
proved by several authors with the different methods 
and generalized to different fields [1-22]. Most of works 
mainly studied the Levinson theorem in the three-dimensional 
space. With the wide interest in lower-dimensional field theories 
recently, the two-dimensional Levinson theorem has been studied 
numerically [18] as well as theoretically [19-25]. 
With respect to the two-dimensional Schr\"{o}dinger equation, 
the version of the Levinson theorem can be read as
$$\eta_{m}(0)=\left\{\begin{array}{ll}
\left(n_{m}+1 \right)\pi~~~~~ &{\rm when}~~m=1~
{\rm and~a~half-bound~state~occurs} \\
n_{m}\pi &{\rm the~remaining~cases}, \end{array} \right. 
\eqno (4) $$

\noindent
where $\eta_{m}(0)$ is the limit of the phase shifts at zero 
momentum for the $m$th partial wave, and $n_{m}$ is the total 
number of bound states with the given angular momentum $m\hbar$. 

Due to the wide interest in lower-dimensional field theory recently, 
it may be worth studying the Levinson theorem in one dimension 
besides the study in two dimensions for completeness. In fact, 
it is a common knowledge that the one-dimensional quantum 
scattering describes many actual physical phenomena to a 
good approximation. For instance, the problem on the tunneling 
times has been discussed in [26]. Furthermore, the one-dimensional 
models are often applied to make the more complex higher-dimensional 
systems tractable. Consequently, it seems reasonable to study the 
one-dimensional Levinson theorem. This will be beneficial for 
understanding both the two-dimensional Levinson theorem and the 
three-dimensional one. Actually, it seems that the direct or implicit 
study of the one-dimensional Levinson theorem [16,27-34] has attracted 
much more attention than that of the two-dimensional one. Nevertheless, 
we attempt to approach this problem by the Sturm-Liouville theorem [35]. 

Generally speaking, there are several methods for studying the 
one-dimensional Levinson theorem for the nonrelativistic particle. 
One is based on the partial-wave analysis method [32, 33]. 
The second relies on the parity-eigenstate representation method 
[16, 34]. The third is to establish the Levinson theorem by 
the Jost function and the $S$ matrix method [29], 
which is essentially based on the orthogonality and completeness 
relation for the eigenfunctions of the total Hamiltonian, as was 
first noticed by Jauch [3].  

The purpose of this paper is to demonstrate the 
one-dimensional Levinson theorem for the Schr\"{o}dinger equation 
by the Sturm-Liouville theorem. We arrive at the final result read as
$$\eta_{+}(0)+\pi/2=n_{+}\pi,~~~~~ \eta_{-}(0)=n_{-}\pi, ~~~~~ 
{\rm for~the~non-critical~case}, $$
$$\eta_{+}(0)=n_{+}\pi,~~~~~ \eta_{-}(0)-\pi/2=n_{-}\pi, ~~~~~ 
{\rm for~the~critical~case}. \eqno (5) $$

\noindent
where the $n_{+}$ and $n_{-}$ denote the number of bound states 
with even parity and odd parity, and the $\eta_{+}(0)$ and $\eta_{-}(0)$ 
denote the phase shift of the scattering states with the 
same parity at zero momentum, respectively. This conclusion
coincides with that shown in [16]. 

It is readily to find from Eq. (5) that the Levinson theorem 
for the odd-parity case in one dimension is the same as that 
for the case $\ell=0$ in three dimensions. However, the 
even-parity case has no counterpart compared to the 
three-dimensional Levinson theorem. This is a very interesting 
feature in one-dimensional Levinson theorem for the 
Schr\"{o}dinger equation. 

This paper is organized as follows. For simplicity, 
we first discuss the cutoff potential case, where 
the potential is vanishing beyond a sufficiently large distance 
$x_{0}$, and leave the discussion for the general case where 
the potential has a tail at infinity in the Sec. V. 
In Sec. II the logarithmic derivative of the wave function of 
the Schr\"{o}dinger equation is chosen as the phase angle [35], 
which is proved to be monotonic with respect to the energy
(the Sturm-Liouville theorem). In Sec. III, according to this 
monotonic property, the number of bound states is shown to be 
related with the the logarithmic derivative of zero energy at 
$x_{0}$ when the potential changes from zero to the given value.  
It will be further shown that the logarithmic derivative of zero 
energy at $x_{0}$ also determines the limit of the phase shifts 
at zero momentum in Sec. IV, which leads to the establishment 
of the one-dimensional Levinson theorem. The critical case, where a 
zero-energy solution occurs, is also analyzed there. 

\section{NOTATIONS AND THE STURM-LIOUVILLE THEOREM}

Throughout this paper the natural units $\hbar=1$ and $2m=1$ 
are employed. Let us consider the one-dimensional 
Schr\"{o}dinger equation with a symmetric potential $V(x)$
$$\frac{d^{2}\psi(x)}{dx^2}+[E-V(x)]\psi(x)=0,~~~~~V(-x)=V(x), $$ 

\noindent
where $E$ denotes the energy of the particle. For simplicity, we 
first discuss the case with a cutoff potential:
$$V(x)=0, ~~~~~{\rm when}~~x\geq x_{0}, \eqno (6) $$

\noindent
where $x_{0}$ is a sufficiently large distance. 
Introduce a parameter $\lambda$ for the potential $V(x)$:
$$V(x, \lambda)=\lambda V(x), \eqno (7) $$

\noindent
where the potential $V(x, \lambda)$ changes from zero to the 
given potential $V(x)$ as $\lambda$ increases from zero to one. 
After introducing the parameter $\lambda$, the one-dimensional
Schr\"{o}dinger equation can be modified as  
$$\displaystyle {\partial^{2} \over \partial x^{2}}
\psi(x, \lambda)+\left[E-V(x, \lambda)\right]\psi(x, \lambda)=0. 
\eqno (8) $$

Since the potential is symmetric, the energy eigenfunctions can 
be combined into those with a definite parity, which satisfy the 
following boundary conditions at the origin:
$$\left. \psi^{(o)}(x, \lambda)\right|_{x=0}=0, ~~~~~~
{\rm for~the~odd-parity~case}, $$
$$\left. \displaystyle {\partial \psi^{(e)}(x, \lambda)
\over \partial x}\right|_{x=0}=0,~~~~~~ 
{\rm for~the~even-parity~case}, \eqno (9) $$

\noindent
Therefore, in the course of studying the one-dimensional Levinson theorem 
we only need to discuss the wavefunction in the range 
$0\leq x<\infty$ with the given parities, even parity case and odd parity one,
respectively. 

Now, we are going to solve Eq. (8) in two ranges $[0,x_{0}]$ and 
$[x_{0},\infty)$, and match two solutions at $x_{0}$. Ignoring the 
effect of the normalization factor, which is irrelevant to our discussion, 
we only need one matching condition at $x_{0}$, which is the 
condition for the logarithmic derivative of the wave function [35]:
$$A(E, \lambda)\equiv \left\{ \displaystyle {1 \over
\psi(x, \lambda) }
\displaystyle {\partial \psi (x, \lambda) \over \partial x}
\right\}_{x=x_{0}{-}} 
=\left\{ \displaystyle {1 \over \psi (x, \lambda) }
\displaystyle {\partial \psi (x, \lambda) \over \partial x}
\right\}_{x=x_{0}{+}}. \eqno (10) $$

According to the condition (9), there exists only one solution 
near the origin. For example, for the free particle ($\lambda=0$), 
the solution to Eq. (8) at the range $[0,x_{0}]$ is real and read as:
$$\psi^{(e)}(x, 0)=\left\{\begin{array}{ll}
\cos(kx)~~~~~&{\rm when}~~E=k^{2}>0\\
\cosh(\kappa x), ~~~~~&{\rm when}~~E=-\kappa^{2}\leq 0, 
\end{array} \right. \eqno (11) $$

\noindent
for the even-parity case, and
$$\psi^{(o)}(x, 0)=\left\{\begin{array}{ll}
\sin(kx)~~~~~&{\rm when}~~E=k^{2}>0\\
\sinh(\kappa x), ~~~~~&{\rm when}~~E=-\kappa^{2}\leq 0, 
\end{array} \right. \eqno (12) $$

\noindent
for the odd-parity case. 

In the range $[x_{0}, \infty)$, we have $V(x)=0$. For $E>0$, 
there exist two oscillatory solutions to Eq. (8) whose combination 
can always satisfy the matching condition (10), so that there
is a continuous spectrum for $E>0$. Assuming that the phase 
shifts $\eta_{\pm}(k,\lambda)$ are zero for the free particles
($\lambda=0$), we have
$$\psi (x, \lambda)=\left\{\begin{array}{ll}
\cos(kx+\eta_{+}(k,\lambda)), ~~~~&{\rm for~the~even-parity~case}\\
\sin(kx+\eta_{-}(k,\lambda)), ~~~~&{\rm for~the~odd-parity~case}, 
\end{array}\right. \eqno (13) $$
$$\eta_{\pm}(k, 0)=0, ~~~~~{\rm when}~~k>0. \eqno (14) $$

We would like to make some remarks here. First, at the first sight, 
the wavefunction in Eq. (13) seems not to have a definite
parity. As a matter of fact, the solutions (13) are only suitable 
in the region $[x_{0},\infty)$. The corresponding solutions in
the region $(-\infty, -x_{0}]$ can be calculated according to the 
parity of the solution. For example, in the odd-parity case, the 
solution in the region $(-\infty, -x_{0}]$ is 
$$-\sin(k|x|+\eta_{-}(k,\lambda))=\sin(kx-\eta_{-}(k,\lambda)). $$

\noindent
Second, the solution (13) for the even-parity case can be rewritten
as
$$\sin(kx+\eta_{+}(k,\lambda)+\pi/2), \eqno (15) $$

\noindent
$\eta_{+}(k,\lambda)+\pi/2$ plays the same role in the even-parity
case as $\eta_{-}(k,\lambda)$ in the odd-parity case. Therefore,
we only need to establish the Levinson theorem for the odd-parity 
case, and the Levinson theorem for the even-parity case can be
obtained by replacing  $\eta_{-}(k,\lambda)$ with
$\eta_{+}(k,\lambda)+\pi/2$.

At last, in the region $[x_{0},\infty)$, the potential $V(x,\lambda)$
is vanishing and does not depend on $\lambda$. However, the
phase shifts $\eta_{\pm}(k,\lambda)$ depend on $\lambda$ through 
the matching condition (10):
$$\tan \eta_{-}(k, \lambda)=-\tan (kx_{0}) ~\displaystyle
{A(E, \lambda)-k\cot (kx_{0})\over A(E, \lambda)+k\tan (kx_{0})}, 
\eqno (16) $$

\noindent
for the odd-parity case, and the similar formula for the 
even-parity case can be obtained by replacing $\eta_{-}(k,\lambda)$ 
with $\eta_{+}(k,\lambda)+\pi/2$.

The phase shifts $\eta_{-}(k, \lambda)$ are determined from Eq. (16) 
up to a multiple of $\pi$ due to the period of the tangent 
function. In our convention (14), the phase shift $\eta_{-}(k, \lambda)$,
$k>0$, changes continuously as $\lambda$ increases from zero to one. 
In other words, the phase shift $\eta_{-}(k, \lambda)$ is 
determined completely in our convention, so is 
$\eta_{+}(k,\lambda)$. For simplicity we define
$$\eta_{\pm}(k)\equiv \eta_{\pm}(k, 1). \eqno (17) $$ 

Since there is only one finite solution at infinity for 
$E\leq 0$, both for the even-parity case and for the odd-parity case:
$$\psi(x, \lambda)=\exp(-\kappa x), ~~~~~
{\rm when}~~x_{0}\leq x < \infty . \eqno (18) $$

\noindent
The solution satisfying the matching condition (10) will not 
always exist for $E\leq 0$.  Except for $E=0$, if and only if 
there exists a solution of energy $E$ satisfying the matching 
condition (10), a bound state appears at this energy. Therefore, 
there is a discrete spectrum for $E \leq 0$. The finite 
solution for $E=0$ is a constant one. It decays not fast enough 
to be square integrable such that it is not a bound state if 
the matching condition (10) is satisfied. 

We now turn to the Sturm-Liouville theorem. Denote
by $\overline{\psi}(x, \lambda)$ the solution to Eq. (8) 
corresponding to the energy $\overline{E}$
$$\displaystyle {\partial^{2} \over \partial x^{2}}
\overline{\psi}(x, \lambda)
+\left[\overline{E}-V(x, \lambda)\right] 
\overline{\psi}(x, \lambda)=0. \eqno (19) $$

\noindent
Multiplying Eq. (8) and Eq. (19) by $\overline{\psi}(x, \lambda)$ and
$\psi(x, \lambda)$, respectively, and calculating their difference, 
we obtain
$$\displaystyle {\partial \over \partial x} \left\{ \psi(x, \lambda)
\displaystyle {\partial \overline{\psi}(x, \lambda) \over \partial x} 
-\overline{\psi}(x, \lambda) \displaystyle {\partial \psi(x, \lambda)
\over  \partial x} \right\}
=-\left(\overline{E}-E\right)
\overline{\psi}(x, \lambda)\psi(x, \lambda).  \eqno (20) $$

\noindent
According to the boundary condition (9), the derivative of the 
wavefunction for the even-parity case and the wavefunction for 
the odd-parity case are vanishing at the origin, respectively. 
Therefore, integrating (20) in the range $0\leq x\leq x_{0}$, we obtain
$$\displaystyle {1 \over \overline{E}-E} \left\{ \psi(x, \lambda)
\displaystyle {\partial \overline{\psi}(x, \lambda) \over \partial x} 
-\overline{\psi}(x) \displaystyle {\partial \psi(x, \lambda) 
\over \partial x} \right\}_{x=x_{0}{-}}
=-\int_{0}^{x_{0}}\overline{\psi}(x, \lambda)\psi(x, \lambda)dx. $$

\noindent
Taking the limit, we arrive at
$$\displaystyle  {\partial A(E, \lambda) \over \partial E}
=\displaystyle  {\partial  \over \partial E}
\left( \displaystyle {1 \over \psi(x, \lambda) }
\displaystyle {\partial \psi (x, \lambda) \over \partial x}
\right)_{x=x_{0}{-}} 
=-\psi(x_{0}, \lambda)^{-2}\int_{0}^{x_{0}}\psi(x, \lambda)^{2}dx \leq 0. 
\eqno (21) $$

\noindent
Similarly, from the boundary condition that when $E< 0$ the 
function $\psi(x, \lambda)$ tends to zero at infinity, and when
$E=0$ the derivative of the function is equal to to zero at 
infinity, we have
$$\displaystyle  {\partial  \over \partial E} \left( \displaystyle 
{1 \over \psi(x, \lambda) }\displaystyle {\partial \psi(x, \lambda) 
\over \partial x} \right)_{x=x_{0}{+}}
=\psi(x_{0}, \lambda)^{-2}\int_{x_{0}}^{\infty}\psi(x, \lambda)^{2}dx>0. 
\eqno (22) $$

\noindent
Therefore, when $E\leq 0$, it is evident that both sides of Eq. (10) 
are monotonic with respect to the energy $E$: as the energy increases, 
the logarithmic derivative of the wave function at $x_{0}{-}$ decreases 
monotonically, but that at $x_{0}{+}$ increases monotonically. 
This is the essence of the Sturm-Liouville theorem. 

\section{THE NUMBER OF BOUND STATES}

In this section we will establish the relation between 
the number of bound states and the logarithmic derivative 
$A(0, \lambda)$ of the wave function at $x=x_{0}{-}$ for zero 
energy when the potential changes, in terms of the monotonic 
property of the logarithmic derivative of the wave function
with respect to the energy $E$. 

For $E \leq 0$, we obtain the logarithmic derivative at $x=x_{0}{+}$ 
from Eq. (18):
$$ \left( \displaystyle {1 \over \psi(x, \lambda) }
\displaystyle {\partial \psi(x, \lambda) \over \partial x}
\right)_{x=x_{0}{+}}
=\left\{\begin{array}{ll} 0&{\rm when}~~E\sim 0 \\
-\kappa \sim -\infty &{\rm when}~~E\longrightarrow  -\infty. 
\end{array} \right. \eqno (23) $$

\noindent
On the other hand, when $\lambda=0$, the logarithmic derivative 
at $x=x_{0}{-}$ can be calculated from Eqs. (11) and (12) for $E\leq 0$: 
$$A(E, 0)=\left(\displaystyle {1 \over \psi(x, 0)}
\displaystyle {\partial \psi(x, 0) \over \partial x}
\right)_{x=x_{0}{-}}=\kappa \tanh(\kappa x_{0})
=\left\{\begin{array}{ll} 0 &{\rm when}~~E\sim 0 \\
\kappa \sim \infty &{\rm when}~~E\longrightarrow - \infty. 
\end{array} \right. \eqno (24) $$

\noindent
for the even-parity case, and
$$A(E, 0)=\left(\displaystyle {1 \over \psi(x, 0)}
\displaystyle {\partial \psi(x, 0) \over \partial x}
\right)_{x=x_{0}{-}}=\kappa \coth(\kappa x_{0})
=\left\{\begin{array}{ll} x_{0}^{-1} &{\rm when}~~E\sim 0 \\
\kappa \sim \infty &{\rm when}~~E\longrightarrow - \infty. 
\end{array} \right. \eqno (25) $$

\noindent
for the odd-parity case. 

It is evident to see from Eqs. (23) and (25) that there is 
no overlap between two variant ranges of two logarithmic 
derivatives for the odd-parity case, namely there is no bound 
state for the free particle in the odd-parity case. 
However, there is one point overlap from Eqs. (23) and (24). 
It means that there is a finite solution at $E=0$ when $\lambda=0$
for the even-parity case. It is nothing but a constant solution. 
This solution is finite but does not decay fast enough at 
infinity to be square integrable. It is not a bound state, 
and called a half bound state. We will discuss the cases with
a half bound state later.

Now, both for the even-parity case and for the odd-parity case,
if $A(0, \lambda)$ decreases across the value zero as $\lambda$ 
increases, an overlap between the variant ranges of two logarithmic 
derivatives of two sides of $x=x_{0}$ appears. Since the logarithmic 
derivative of the wave function at $x_{0}{-}$ decreases monotonically 
as the energy increases, and that at $x_{0}{+}$ increases monotonically, 
the overlap means that there must exist one and only one energy 
for which the matching condition (10) is satisfied, that is, a 
bound state appears. From the viewpoint of node theory, when 
$A(0, \lambda)$ decreases across the value zero, a node for the 
zero-energy solution to the Schr\"{o}dinger equation comes 
inwards from the infinity, namely, a scattering state changes to 
a bound state. 

As $\lambda$ increases again, $A(0, \lambda)$ may decreases to 
$-\infty$, jumps to $\infty$, and then decreases again across 
the value zero, so that another overlap occurs and another bound 
state appears. Note that when the zero point in the zero-energy 
solution $\psi(x, \lambda)$ comes to $x=x_{0}$, $A(0, \lambda)$ goes 
to infinity. It is not a singularity. 

Each time $A(0, \lambda)$ decreases across the value zero, 
a new overlap between the variant ranges of two logarithmic 
derivatives appears such that a scattering state changes to 
a bound state. At the same time, a new node comes inwards
from infinity in the zero-energy solution to the Schr\"{o}dinger 
equation. Conversely, each time $A(0, \lambda)$ increases 
across the value zero, an overlap between those 
two variant ranges disappears so that a bound state changes 
back to a scattering state, and simultaneously, a node goes
outward and disappears in the zero-energy solution. The 
number of bound states $n_{\pm}$ is equal to the times that 
$A(0, \lambda)$ decreases across the value zero as $\lambda$ 
increases from zero to one, subtracted by the times that 
$A(0, \lambda)$ increases across the value zero. It is also 
equal to the number of nodes in the zero-energy solution. 
In the next section we will show that this number is nothing but 
the phase shift at zero momentum divided by $\pi$, i.e., 
$\eta_{-}(0)/\pi$ or $\eta_{+}(0)/\pi+1/2$. 

We should pay some attention to the critical case where
$A(0, 1)=0$. A finite zero-energy solution $\psi(x,1)=c$
at $[x_{0},\infty)$ will satisfy the matching condition (10) 
with the zero $A(0, 1)$. Note that when $A(0,1)=0$, the wave 
function at $x_{0}-$, $\psi(x_{0},1)$, must be nonvanishing 
for the non-trivial solution. The constant $c$ is nothing but 
the non-vanishing value $\psi(x_{0},1)$. The constant solution is 
not square integrable so that it is not a bound state, and called a 
half bound state. As $\lambda$ increases from a number near 
and smaller than one and finally to reach one, if $A(0, \lambda)$ 
decreases and finally reaches the value zero, a scattering state
becomes a half bound state, and no new bound state appears. 
Conversely, as $\lambda$ increases to reach one, if $A(0, \lambda)$ 
increases and finally reaches the value zero, a bound state
becomes a half bound state, namely, a bound state disappears.
This conclusion holds for both the even-parity case and the 
odd-parity case.

\section{LEVINSON'S THEOREM}

When $\lambda=0$, the phase shifts $\eta_{\pm}(k, 0)$ are defined
to be zero. As $\lambda$ increases from zero to one, and 
$\eta_{\pm}(k, 0)$ for $k>0$ change continuously. 

For the odd-parity case, the phase shift $\eta_{-}(k, \lambda)$ 
is calculated by Eq. (16). It is easy to see that the phase shift 
$\eta_{\pm}(k, \lambda)$ increases monotonically as the 
logarithmic derivative $A(E, \lambda)$ decreases:
$$\left. \displaystyle {\partial \eta_{-}(k, \lambda) \over 
\partial A(E, \lambda)}\right|_{k}
=\displaystyle {-k\cos^{2}\eta_{-}(k, \lambda) \over
\left\{A\cos(kx)+k\sin(kx)\right\}^{2} } \leq 0, \eqno (26) $$

The phase shift $\eta_{-}(0, \lambda)$ is the limit of the phase 
shift $\eta_{-}(k, \lambda)$ as $k$ tends to zero. Therefore, we 
are only interested in the phase shift $\eta_{-}(k, \lambda)$ at 
a sufficiently small momentum $k$, $k\ll 1/x_{0}$. For the small 
momentum we obtain from Eq. (16)
$$\tan\eta_{-}(k, \lambda) \sim -\left(kx_{0}\right) ~\displaystyle
{A(0, \lambda)-c^{2}k^{2}-x_{0}^{-1}+k^{2}x_{0}/3 \over 
A(0, \lambda)-c^{2}k^{2}+k^{2}x_{0}}, \eqno (27) $$

\noindent
where the expansion of $A(E, \lambda)$ for the small $k$ is
used
$$A(E, \lambda)\sim A(0, \lambda)-c^{2}k^{2}, ~~~~~c^{2}\geq 0, \eqno (28) $$

\noindent
which is calculated from the sturm-Liouville theorem (21).
In both the numerator and the denominator of Eq. (27) we 
included the next leading term, which is only useful for the
critical cases where the leading terms are canceled each other. 

First, it can be seen from Eq. (27) that, except for the special
point where $A(0, \lambda)=0$, $\tan \eta_{-}(k, \lambda)$ 
tends to zero as $k$ goes to zero, namely, $\eta_{-}(0, \lambda)$ 
is always equal to the multiple of $\pi$ except for zero $A(0, \lambda)$.
In other words, if the phase shift $\eta_{-}(k, \lambda)$ 
for a sufficiently small $k$ is expressed as a positive or negative 
acute angle plus $n\pi$, its limit $\eta_{-}(0, \lambda)$ is equal to 
$n\pi$, where $n$ is an integer. It means that $\eta_{-}(0, \lambda)$ 
changes discontinuously. When $A(0, \lambda)=0$, the limit 
$\eta_{-}(0, \lambda)$ of the phase shift $\eta_{-}(k, \lambda)$
is equal to $(n+1/2)\pi$. It is not important for our discussion
except for $A(0, 1)=0$, which we call the critical case and will
discuss the critical case later.

Second, for a sufficiently small $k$, if $A(E, \lambda)$ 
decreases as $\lambda$ increases, $\eta_{-}(k, \lambda)$ 
increases monotonically. Assume that in the variant process
$A(E, \lambda)$ may decreases through the value zero, but does 
not stop at this value. As $A(E, \lambda)$ decreases, each 
times $\tan \eta_{-}(k, \lambda)$ for the sufficiently 
small $k$ changes sign from positive to negative, 
$\eta_{-}(0, \lambda)$ jumps by $\pi$. However, each 
times $\tan \eta_{-}(k, \lambda)$ changes sign from negative to 
positive, $\eta_{-}(0, \lambda)$ remains invariant. Conversely, 
if $A(E, \lambda)$ increases as $\lambda$ increases, 
$\eta_{-}(k, \lambda)$ decreases monotonically. As $A(E, \lambda)$ 
increases, each time $\tan \eta_{-}(k, \lambda)$ changes sign from 
negative to positive, $\eta_{-}(0, \lambda)$ jumps by $-\pi$, and each 
time $\tan \eta_{-}(k, \lambda)$ changes sign from positive 
to negative, $\eta_{-}(0, \lambda)$ remains invariant.  

Third, as $\lambda$ increases from zero to one, $V(x, \lambda)$ 
changes from zero to the given potential $V(x)$ continuously.
Each time the $A(0, \lambda)$ decreases from near and larger than 
the value zero to smaller than that value, the denominator in 
Eq. (27) changes sign from positive to negative and the remaining 
factor remains positive, such that the phase shift at zero momentum 
$\eta_{-}(0, \lambda)$ jumps by $\pi$. Conversely, each time the 
$A(0, \lambda)$ increases across the value zero, the phase shift 
at zero momentum $\eta_{-}(0, \lambda)$ jumps by $-\pi$. Each time 
the $A(0, \lambda)$ decreases from near and larger than 
the value $x_{0}^{-1}$ to smaller than that value, the numerator in 
Eq. (27) changes sign from positive to negative, but the remaining 
factor remains negative, such that the phase shift at zero momentum 
$\eta_{-}(0, \lambda)$ does not jump. Conversely, each time the 
$A(0, \lambda)$ increases across the value $x_{0}^{-1}$, the phase shift 
at zero momentum $\eta_{-}(0, \lambda)$ does not jump, either.  

Therefore, the phase shift $\eta_{-}(0)/\pi$ is just equal to the
times $A(0, \lambda)$ decreases across the value zero 
as $\lambda$ increases from zero to one, subtracted by the times 
$A(0, \lambda)$ increases across that value. 
As discussed in the previous section, we have proved that the 
difference of the two times is nothing but the number of bound 
states $n_{-}$, namely, for the non-critical cases, the Levinson 
theorem for the one-dimensional Schr\"{o}dinger equation 
in the odd-parity case is 
$$\eta_{-}(0)=n_{-}\pi. \eqno (29) $$

Fourth, we now turn to discuss the critical case where the 
logarithmic derivative $A(0, 1)$ ($\lambda=1$) is equal to 
zero. In the critical case, the constant solution $\psi(x)=c$
($c\neq 0$) in the range $[x_{0},\infty)$ for zero energy will match 
this $A(0, 1)$ at $x_{0}$. In the critical case, it is obvious 
that there exists a half-bound state both for the even-parity 
case and for the odd-parity case. A half-bound state 
is not a bound state, because its wave function is finite 
but not square integrable. As $\lambda$ increases from a number 
near and less than one and finally reaches one, if the logarithmic 
derivative $A(0, \lambda)$ decreases and finally reaches, but
not across, the value zero, according to the discussion in the
previous section, a scattering state becomes a half
bound state when $\lambda=1$. On the other hand, the denominator
in Eq. (27) is proportional to $k^{2}$ such that $\tan \eta_{-}(k,1)$
tends to infinity. Namely, the phase shift $\eta_{-}(0,1)$ jumps
by $\pi/2$. Therefore, for the critical case the Levinson theorem becomes
$$\eta_{-}(0)-\pi/2=n_{-}\pi. \eqno (30) $$

Conversely, as $\lambda$ increases and reaches one, if the 
logarithmic derivative $A(0, \lambda)$ increases and finally 
reaches the value zero, a bound state becomes a half bound 
state when $\lambda=1$, and the phase shift $\eta_{-}(0,1)$ 
jumps by $-\pi/2$. In this situation, the Levinson theorem (30)
still holds. 

At last, for the even-parity case, the only change is to replace
the phase shift $\eta_{-}(0)$ with the phase shift 
$\eta_{+}(0)+\pi/2$. Therefore, the Levinson 
theorem for the one-dimensional Schr\"{o}dinger equation 
in the even-parity case is 
$$\eta_{+}(0)+\pi/2=n_{+}\pi,~~~~~{\rm for~the~non-critical~cases}, $$
$$\eta_{+}(0)=n_{+}\pi,~~~~~{\rm for~the~critical~cases}. \eqno (31) $$

Note that for the free particle in the even-parity case, 
there is a half bound state at $E=0$. It is the critical
case where $\eta_{+}(0)=0$ and $n_{+}=0$. Combining Eqs. 
(29-31), we obtain the Levinson theorem for the 
one-dimensional Schr\"{o}dinger equation as Eq. (5).

\section{DISCUSSIONS}

Now, we discuss the general case where the potential $V(x)$ 
has a tail at $x\geq x_{0}$. First, we assume that
$$V(x)=bx^{-2},~~~~~x\geq x_{0}. \eqno (32) $$

\noindent
It is obvious that when $b<-1/4$ there is an infinite number
of bound states for the Schr\"{o}dinger equation (8) such
that the Levinson theorem (5) is violated. When $b\geq -1/4$, 
let
$$j(j+1)=b,~~~~~j=-1/2+(b+1/4)^{1/2}\geq -1/2. \eqno (33) $$

\noindent
The Schr\"{o}dinger equation (8) becomes the same as the
radial equation in three dimensions except that the phase shift
is $\eta_{-}(k,\lambda)-j\pi/2$ now. Repeating the proof in our previous 
paper [6], we obtain the modified Levinson theorem for the
Schr\"{o}dinger equation (8) with the potential (32)
in the non-critical cases:
$$\eta_{-}(0)-j\pi/2=n_{-}\pi,~~~~~
\eta_{+}(0)+(1-j)\pi/2=n_{+}\pi. \eqno (34) $$

\noindent
In other words, the Levinson theorem (5) is violated. It is
obvious that the Levinson theorem will be violated more seriously
if the potential tail decays at infinity slower than the potential 
tail (32). On the other hand, if the potential tail decays at 
infinity faster than the potential tail (32), for an arbitrarily 
given small positive number $\epsilon$, there always exists a 
larger enough number $x_{0}$ such that
$$(-\epsilon)(-\epsilon+1)x^{-2} < V(x) < \epsilon(\epsilon+1)x^{-2},
~~~~~x\geq x_{0}. \eqno (35) $$ 

\noindent
Since $\epsilon$ is arbitrarily small, no modification is needed 
to the Levinson theorem (5). 

In conclusion, we establish the one-dimensional Levinson 
theorem (5) for the Schr\"{o}dinger equation in one dimension
with the potential satisfying
$$V(-x)=V(x),~~~~~\displaystyle \lim_{x\rightarrow \infty}x^{2}V(x)=0.
\eqno (36) $$

\vspace{6mm}
{\bf ACKNOWLEDGMENTS}. This work was supported by the National 
Natural Science Foundation of China and Grant No. LWTZ-1298 
from the Chinese Academy of Sciences.

\end{document}